\documentclass[twocolumn,showpacs,preprintnumbers,amsmath,amssymb]{revtex4}

\usepackage{graphicx}
\usepackage{dcolumn}
\usepackage{bm}

\usepackage{amsmath}

\begin{document}

\title{Persistent Currents in Small, Imperfect Hubbard Rings} 
\author{P. Koskinen and M. Manninen} 
\affiliation{Department of Physics, 40014 University of Jyv\"askyl\"a, Finland} 
\date{\today} 
 
\begin{abstract} 
 We have done a study with small, imperfect Hubbard rings with 
exact diagonalization. The results for few-electron rings show, 
that the imperfection, whether localized or not, nearly always 
decrease, but can also \emph{increase} the persistent current, 
depending on the character of the imperfection and the on-site interaction.  
The calculations are  
generally in agreement with more specialized studies. In most cases 
the electron spin plays an important role. 
\end{abstract} 
  

\maketitle

\section{Introduction} 
The regime of experimental studies in small semiconductor 
hetero\-struc\-tures has gone from mesoscopic to nanoscopic within a 
few years. Especially electron traps have been 
under intensive study because of the discrete nature of energy 
levels and the resulting analogy with real atoms. The trapping potential has varied 
between parabolic (quantum dot) \cite{tarucha96} and ring-like (quantum ring, QR) 
\cite{lorke99, fuhrer01, keyser02}. While quantum dots 
fascinate because of the similarity with real atoms, rings combine 
this similarity with an always captivating ring-like geometry.

Rings are often studied with a relation to the 
persistent current (PC) \cite{buttiker83}, an equilibrium current that arises when an 
Aharonov-Bohm flux is piercing the ring \cite{aharonov59}. 
Few-electron nanoscopic rings introduce many-body effects not 
observable in mesoscopic rings, such as the fractional 
$\phi_0/N$-periodicity of the persistent current 
\cite{kusmartsev95}. These systems have been studied 
theoretically, both in the single-particle 
\cite{planelles01,llorens01} as well as in the many-body picture 
\cite{chakraborty94,halonen96,viefers00,koskinen01}. With continuum models, 
particularly in the analytical approach, it is laborious to 
introduce imperfections to a perfect ring, especially when non-perturbative 
treatment is required.

In addition to continuum models, lattice models have been applied to QRs 
\cite{abraham93,cheung88,bouzerar94,yu92}. If interactions are 
taken into account, the Hubbard model \cite{hubbard63} is probably 
the most investigated model. Now, apart from being a toy model 
of mathematical physics \cite{lieb68}, the 
purpose of this paper is to show, that a slightly generalized 
Hubbard model can also be an extremely valuable tool in getting the 
first crude idea of what is going on in small, numerically exactly 
diagonalizable systems. With the Hubbard model and exact 
diagonalization it is easy to introduce the effects, including for example an impurity, 
disorder, magnetic- and electric fields and external leads, that are much 
harder to include with other models. Here one 
does not need to limit to a certain region in the parameter space but 
can explore all the possible values of the parameters and we get an illustrative 
representation of the results.

The calculations show, that imperfections mainly decreases the persistent 
current. The interaction, however, can introduce coupling of the 
two spin-currents and result in increase of the PC in certain situations, 
especially when the imperfection is of localizing nature. We conclude that 
spinless models cannot show these effects.

\section{The Model} 
We use the Hubbard Hamiltonian with a pure 
Aharonov-Bohm flux without a Zeeman term. In the presence of a 
vector potential the hopping integral is 
modified by a phase factor \cite{peierls33} and the Hamiltonian reads 
\begin{eqnarray} 
\label{ham} 
\begin{array}{c}
H = -\sum_{ i,j,  \sigma} t_{ij} ( e^{-ieA_{ij}/\hbar}c_{i 
\sigma}^\dagger c_{j \sigma} + H.c.) \\
\\
+ U\sum_i \hat{n}_{i 
\uparrow} \hat{n}_{i \downarrow} + \sum_{i, \sigma} \epsilon_i 
\hat{n}_{i \sigma}, 
\end{array}
\end{eqnarray} 
where we have generalized for site-dependent hopping integral 
as well as for one-body on-site energies $\epsilon_i$. $A_{ij}$ 
represents the portion of the Aharonov-Bohm flux covered by the 
hopping $i\leftrightarrow j$, so that we could write the phase factor as 
$exp[-i (2 \pi/L) (\phi/ \phi_0) ]$, where $L$ is the number of 
sites in the ring, $\phi$ is the magnetic flux and $\phi_0=h/e$ 
the flux quantum. Energy scale is fixed by setting all $t_{ij}=1$ 
unless otherwise stated. 
Figure 1a shows examples of Hubbard rings considered in this research.

\begin{figure}
\includegraphics{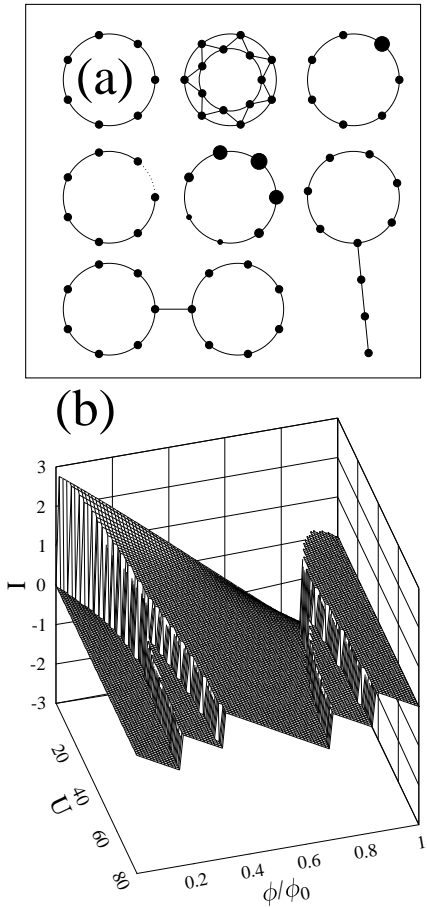}
\caption{(a) Structures of some of the Hubbard rings used in the calculations.
The size of the point represents the site energy. The dashed line 
represents a weaker (or stronger) hopping propability.
(b) Persistent current of a perfect ring with $L=7$ and $N_\uparrow=N_\downarrow=2$
as a function of the interaction $U$ and the flux  $\phi$. 
Note the periodicity change from
$\phi_0$ to $\phi_0/N_e=\phi_0/4$ as the interaction increases.}
\label{fig1}
\end{figure}

The dimension of $H$ is $\binom{L}{N_\uparrow} 
\binom{L}{N_\downarrow}$, where $N_\sigma$ is the number of 
spin-$\sigma$ -electrons, growing strongly with $L$ and 
$N_\sigma$, limiting the system size. The diagonalization is done 
with standard library routines and the numerics do not involve 
severe difficulties.

The persistent current of an eigenstate $\psi_m$ in a single 
ring is usually viewed via the expression $I_m(\phi)=-\partial 
E_m/\partial \phi$ \cite{byers61,bloch70}. In the presence of multiple 
rings and different currents, however, we have to use the 
current operator. From the relation above and from the 
Feynman-Hellmann theorem we obtain \cite{scalapino92} 
\begin{equation} 
\label{curr} 
\hat{j}_{kl \sigma} = \frac{4 \pi t_{kl}}{\phi_0} \mathrm{Im} [ e^{-i (2 
\pi/L) (\phi/\phi_0)}c_{k \sigma}^\dagger c_{l \sigma}] 
\end{equation} 
for the spin-$\sigma$ -electron current operator between the (irrelevant) sites $k$ and 
$l$. The persistent current with the definition 
$I(\phi)=-\partial E/\partial \phi$ is the sum of the two 
different spin currents. Using this operator is beneficial also 
because we obtain the current directly by taking the expectation 
value of the operator (\ref{curr}) in a given state.

The persistent current is a periodic function of $\phi$ with periodicities 
$\phi_0$, $\phi_0/2$ and $\phi_0/N_e$, increasing with the interaction, as 
shown in Fig.1b. The periodicity has been studied before 
\cite{koskinen02,viefers03,kusmartsev95}, and the purpose of Fig.1b is just to show 
that it makes sense to characterize the persistent current by fixing $\phi$ as long as it does not 
contain any discontinuities. Like many other authors \cite{viefers00,abraham93}, 
we have used the value $\phi=0.25 \phi_0$ and the ground state persistent current 
throughout the paper.

It is suggestive, that in the Hubbard model the 
quasi-one-dimensionality can be mimicked by adjusting the interaction term: 
A Hubbard ring with multiple channels (Fig.1a) and $U\gg 1$ is mimicked by a single 
channel and a relatively small $U$ \cite{viefers03}. Thus the Hubbard $U$ in these ring-like structures 
can be read either as an interaction as such or as a measure of the 
one-dimensionality of the ring.

\section{Results and Discussion} 
 
The effect of a single impurity on PC has been studied by several authors 
\cite{viefers00, chakraborty95, monozon03}. Fig.2a shows PC influenced 
by an on-site energy $\epsilon_1$ and the interaction. 
With zero impurity, the current decreases when the interaction $U$ (or one-dimensionality) increases.  
This happens because with increasing $U$ the Hamiltonian (\ref{ham}) 
prevents electrons moving independently and eventually only rotation as a rigid rotor is possible. 
These observations are in agreement with more 
specialized papers \cite{viefers00,niemela96}.  
 
\begin{figure*}
\includegraphics{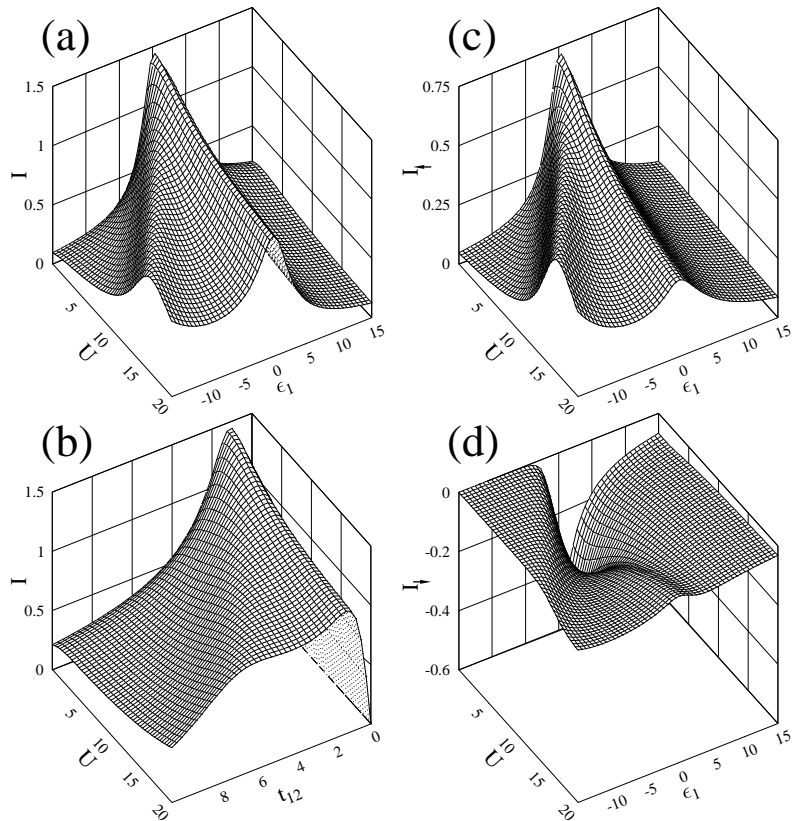}
\caption{Effect of a single impurity on the persistent current
at a fixed value of flux ($\phi=0.25\phi_0$). In (a) the impurity is 
described with a onsite energy $\epsilon_1$, in (b) with a hopping
parameter $t_{12}$ ($L=7$, $N_\uparrow=N_\downarrow=2$).
(c) and (d) show separately the spin-up and spin-down currents 
in the case of three electrons 
($L=7$, $N_\uparrow=2$, $N_\downarrow=1$) and a single impurity
described with $\epsilon_1$.}
\label{fig2}
\end{figure*}

The current always decreases monotonously with increasing positive $\epsilon_1$.  
Investigating at the Fig.2a more carefully with a fixed 
positive impurity, one can see that there is a slight increase of 
the current with the interaction. This result, 
confirming earlier calculations \cite{kamal95,mullergroeling94}, 
tells that a greater interaction makes the electron system 
more correlated, pushing the impurity-affected electrons 
more effectively. Because there are observations 
contradicting with this result \cite{chakraborty95}, 
saying that in the presence of impurity, the 
interactions suppress PC even further, we hasten to add that 
with $\epsilon_1>0$ the current with large enough 
interaction is slightly smaller than the current without interaction. 
Note that in the case of, for example, the Coulomb potential, the interaction 
is either on or off, and this kind of gradual increase in the interaction 
strength is usually not considered. However, as discussed above,
the increase of electron-electron interaction 
corresponds to decrease of the width of the quasi-one-dimensional ring.

With often disregarded attractive impurity and fixed $U$ one can see a local maximum of $I$  
with specific $\epsilon_1$. 
Here the impurity localizes electrons, and if the depth 
of the impurity potential is equal to the interaction, $U+\epsilon_1=0$, the electrons 
with the opposite spin as an effect do not feel any extra on-site energy and thus 
pass the impurity unaffected. Either increasing or 
decreasing $U$ (or $\epsilon_1$) makes the effective on-site energy repulsive or attractive, 
decreasing $I$. Away from this region of local maximum (especially with $U=0$) the attractive 
impurity decreases the current more effectively than the repulsive one.

A similar effect is seen in Fig.2b, where the 
"impurity" is now a different hopping integral $t_{12}$. 
The current goes to zero with $t_{12}$ as it should, but 
surprisingly, it has a maximum near $t_{12}\sim 1$, above which it 
decreases. The physics is the same as with attractive impurity, because 
the large negative kinetic energy of the strong link localizes an electron 
pair to the corresponding sites, creating a similar kind of blocking 
effect; notice that the form of Fig.2b with $t_{12}>1$ is the same as 
in Fig.2a with $\epsilon_1<0$.

Because we have an equal number of electrons with the opposite spin, 
the different spin-currents, by symmetry, are the same. But if 
$N_\uparrow \neq N_\downarrow$, they are in the \emph{opposite} 
direction. This can be seen in Figs.2c and 2d, which show spin-up 
and spin-down-currents as a function of $U$ and the single 
impurity strength for $N_\uparrow=2$ and $N_\downarrow=1$. 
The opposite signs and relative magnitudes of PC with 
zero interaction and impurity, i.e. 
$I_\uparrow(0,0)/I_\downarrow(0,0)$, can in fact be easily explained also with
{\it non-interacting} electrons in a continuous, strictly one-dimensional ring. 
The effect of interaction 
is to `grab' the spin-down electron to move to the same direction 
as the spin-up electrons, leading eventually to the rigid rotation of the 
whole electron system. 
However, to the total current ($I_\uparrow+I_\downarrow$) 
the impurity has the same effect in the three-electron case as 
in the four-electron case. With other odd numbers of electrons the
situation is similar to that of the three-electron case.

\begin{figure}
\includegraphics{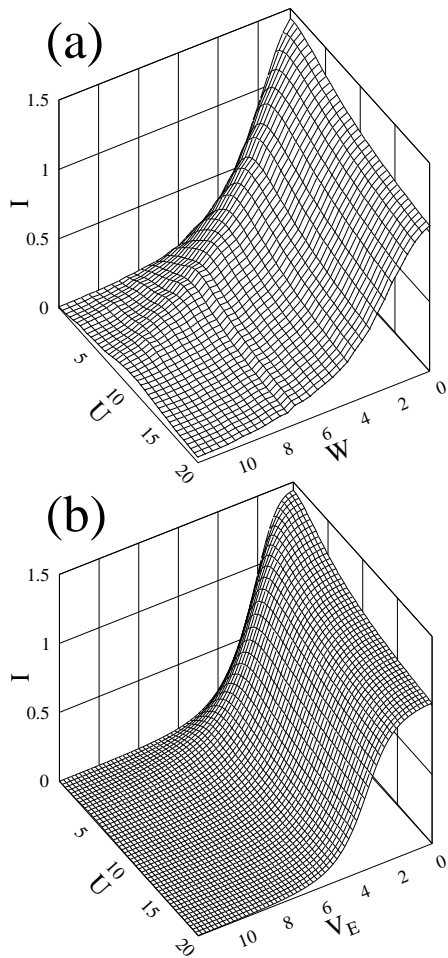}
\caption{Effect of random on-site energies (a) and electric filed (b) 
on the persistent current at $\phi=0.25\phi_0$
($L=7$, $N_\uparrow=N_\downarrow=2$).
$W$ is a measure of the disorder of the onsite energies and 
$V_E$ is the strength of the electric field parallel to the plane of the ring.}
\label{fig3}
\end{figure}

A random external potential at the ring is often called disorder
and its effect on PC and the energy spectra is largely 
studied subject mainly due to the relevance in experimental samples 
\cite{abraham93,bouzerar94,cheung88}. Fig.3a shows an ensemble  
averaged $I(U,W)$, where $W$ is the measure of the disorder, 
defined as $W=max(\{\epsilon_i\})-min(\{\epsilon_i\})$, and 
$\{\epsilon_i\}$ are random on-site energies
(the enesemble consisted of 60 rings).
The current decreases 
monotonously with $W$, but the role of the interactions is again not 
trivial, since a given disorder strength gives the current a maximum with $U\sim W$.  
The increase of PC with $U(\ll 1)$ has 
been confirmed 
by other authors \cite{kamal95,mullergroeling94}. 
Physically the maximum can again be 
explained by the effective potential-smoothening 
that originates from the disorder-localized electrons and their repulsive 
interaction towards the electrons of opposite spin. 
The impact of interactions can be different in spinless models \cite{bouzerar94}(where inter-site 
interaction is required, e.g. a Coulomb-like interaction term \cite{abraham93}). 
But at this point we want to stress that if the spin-currents, 
potentially even of opposite sign, 
are present, then the effect of interaction, by coupling the 
two currents as described above, is quite subtle and differs from 
the effect in spinless models.

By comparing carefully Figs.3a and 3b, the resemblance is obvious.  
Fig.3b shows the same plot as Fig.3a but 
now with disorder strength $W$ replaced by the strength $V_E$ of an electric 
field parallel to the plane of the quantum ring. The physics remain 
the same: while the field `inclines' the ring so that some electrons 
tend to roll down to the other end of the ring, the interaction 
smoothens the way to the electrons of the opposite spin, creating 
an optimum interaction for given $V_E$. On the other hand, for a given $U$ an increasing 
electric field destroys the Aharonov-Bohm oscillations, 
since the wave-function does not extent throughout the whole ring 
\cite{barticevic02}. 
 
\begin{figure}
\includegraphics{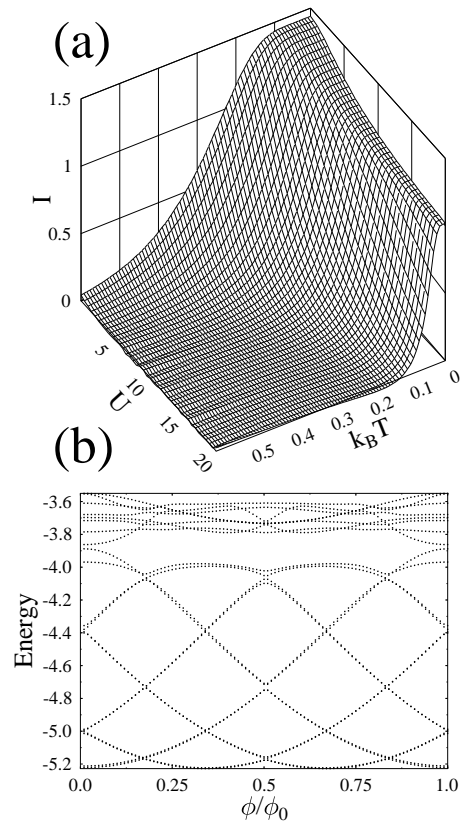}
\caption{(a) Temperature dependence of the persistent current in a perfect ring
with with $L=7$ and $N_\uparrow=N_\downarrow=2$
($k_B T$ measured in units of $t$). 
(b) The energy spectrum $E_i(\phi)$ of a ring with six sites and a stub with two sites
(in Fig. 1a is shown a related system with a ring of seven sites and a stub of
three sites). The system has four electrons ($N_\uparrow=N_\downarrow=2$)
and $U=1000$. The low-energy spectrum is nearly identical with a spectrum of three 
electrons in a pure ring without the stub.}
\label{fig4}
\end{figure}

Finite temperature behaviour of the persistent current is more 
intuitive, as seen in Fig.4a. The current decreases monotonously 
with $U$ at all temperatures, and with a fixed $U$  
the current is almost constant for small enough $T$, until it decreases essentially to 
zero in the temperature range that is given by the energy gap 
between the ground state and the excited states. This is because 
the directions of the currents of the exited states are frequently 
opposite to the current of the ground state. The observations 
agree with the non-interacting picture of Refs.\cite{buttiker85, 
cheung88}.

As is the temperature, so are the external leads an inevitable 
imperfection in externally tuneable quantum rings \cite{fuhrer01,keyser02}. 
With a lattice model one could imagine modelling these 
with an attractive on-site energy, larger hopping integral or 
maybe with a smaller interaction term that is due to the larger space 
available in the vicinity of the leads. The ring-stub -system, a 
schematic of which is shown in Fig.1a, has been studied by 
continuum models \cite{sreeram96, moskalets00,deo00}. It was 
found, that the stub can create standing waves which are not 
affected by the magnetic flux piercing the ring. With our 
approach, including now also the many-body effects, it is found 
that with large enough interaction, the stub of definite length 
can localize an integer number $N_{stub}$ of electrons, and as a 
result the remaining electrons in the ring exhibit normal 
AB-oscillations with $N_e-N_{stub}$ electrons, \emph{as if} the 
stub would be absent. This is shown in Fig.4b, where we have a system 
of four electrons, but the low-energy spectrum is essentially the 
same as the pure-ring spectrum with just three electrons. If the 
stub-length would be increased by one site, the standing wave 
would not have a node anymore at the stub-ring interface and the 
localized electron would start interfering with the electrons in 
the ring, suppressing the current.

If the stub is connected to another quantum ring, as depicted in 
Fig.1a, we get coupled quantum rings, which could be 
experimentally realistic in a dense ensemble of 
self-assembled quantum rings \cite{lorke99}. Studied with 
non-interacting continuum electrons \cite{pareek96}, the relative 
directions of PC in coupled rings of different radius were found to 
depend on the magnetic flux. If we define a reference ring with 
length $L_r$ to enclose a flux $\phi_r$,  a ring with 
length $L$ then encloses a flux $\phi=(L/L_r)^2 \phi_r$, and we can have 
rings of different sizes by applying the corresponding fluxes to 
the phase factor of the hopping integral in the Hamiltonian (\ref{ham}). It is 
found that the currents in the rings of length $L_1$ and $L_2$ 
indeed run in the opposite direction if e.g. the flux through the 
first ring is less and the flux through the second ring is 
more than integer multiple of a flux quantum \cite{oddness}.  
Though we have two independent currents, it turns out that the derivative of the total energy 
resembles the sum of the currents very closely, and thus may 
lead to large periodicities if $L_1/L_2$ differs only slightly from 
unity.

Most of the figures show results for rings with $L=7$ and $N_e=4$, but it is important 
to point out that also other rings were investigated, giving similar 
results. We want to defend this choice by the notion that 
in the limit of large $U$, the Hubbard model becomes 
the Heisenberg model with an effective coupling constant $J_{eff}$ 
\cite{yu92,viefers03}, which, for large $L$, scales as 
$J_{eff} \propto L^{-3}$. This, 
together with our calculations, shows that the energy separations -and 
consequently the persistent current- decreases with $L$. This implies that 
the general trends depicted 
in the figures tend to diminish as the number of empty lattice sites 
is increased. Furthermore, these few-electrons systems, because of the  
\emph{local} interaction, the model does 
not necessarily compare to real continuum models better if 
$L$ is increased.

\section{Conclusions} 
To conclude, we have studied small, imperfect Hubbard rings with exact 
diagonalization. The results show, that the imperfection, 
whether localized or not, almost always decrease the persistent 
current. The interaction, however, can introduce coupling of the 
two spin-currents, make the effective potential felt by opposite-spin 
electrons smoother and result in increase of PC in certain situations, 
especially when the imperfection is of localizing nature. We note that 
these effects cannot be seen with spinless electrons. 
The decrease in PC as a function of impurity potential starts with zero slope. 
The current decreases 
with temperature monotonously in a scale given by the energy gap 
above the ground state.

By comparing with other studies, it is shown that even these 
very small rings give the same physics and phenomena 
that is obtained with more involved Hubbard-model 
calculations, including e.g. Bethe-Ansatz -calculations.  
Furthermore, many realistic investigations with continuum models   
leave the phenomena obtained with this approach mainly intact. 
 
\section{Acknowledgments}

We would like to thank Susanne Viefers and Prosenjit Singha Deo for
valuable discussions.
One of us (PK) acknowledges the V\"ais\"al\"a foundation 
for financial support. 
This work has been supported by the Academy of Finland under the Finnish 
Centre of Excellence Programme 2000-2005 (Project No. 44875, Nuclear and 
Condensed Matter Programme at JYFL).


\begin{thebibliography}{mullergroeling94}
\bibitem{oddness} The persistent current is an odd function of the flux.
\end{thebibliography}
\end{document}